\documentclass[useAMS,usenatbib]{mn2e}

\usepackage{graphicx}
\usepackage{color}
\usepackage{amsmath, amssymb}
\usepackage{natbib}
\usepackage{ulem}
\newcommand{\rlight}{r_{\rm L}}

\title[Thermal synchrotron radiation from DTM]
{Thermal synchrotron radiation from RRMHD simulations of the double tearing mode reconnection - Application to the Crab flares}

\author[Takamoto, P\'etri, \& Baty]{M. Takamoto,$^{1,2}$\thanks{E-mail:
mtakamoto@eps.s.u-tokyo.ac.jp}
J. P\'etri$^3$\thanks{E-mail:
jerome.petri@astro.unistra.fr}
and
H. Baty$^3$\thanks{E-mail:
hubert.baty@astro.unistra.fr}
\\
$^{1}$Max-Planck-Institut f\"ur Kernphysik, Postfach 103980, 69029 Heidelberg, Germany\\
$^{2}$Department of Earth and Planetary Science, University of Tokyo, Tokyo 113-0033, Japan \\
$^{3}${Observatoire astronomique de Strasbourg, Universit\'e de Strasbourg, CNRS, UMR 7550, 11 rue de l'universit\'e, F-67000 Strasbourg, France.}
}
\begin{document}

\date{Accepted 2015 December 15. Received 2015 December 14; in original form 2015 October 11}

\pagerange{\pageref{firstpage}--\pageref{lastpage}} \pubyear{2015}

\voffset=-0.8in

\maketitle

\label{firstpage}

\begin{abstract}
We study the magneto-hydrodynamic tearing instability occurring in a double current sheet configuration when a guide field is present. This is investigated by means of resistive relativistic magneto-hydrodynamic (RRMHD) simulations. Following the dynamics of the double tearing mode (DTM), we are able to compute synthetic synchrotron spectra in the explosive reconnection phase. The pulsar striped wind model represents a site where such current sheets are formed, including a guide field. The variability of the Crab nebula/pulsar system, seen as flares, can be therefore naturally explained by the DTM explosive phase in the striped wind. 
Our results indicate that 
the Crab GeV flare can be explained by the double tearing mode in the striped wind region 
if the magnetization parameter $\sigma$ is around $10^5$. 
\end{abstract}


\section{Introduction}

In a recent study \citep[also referenced as Paper1 below]{2013MNRAS.436L..20B}, a model has been proposed to address the origin of the strong flares observed in X-rays and gamma-rays  from pulsars and magnetars environments \citep{2005Natur.434.1107P,2011ApJ...741L...5S,2014ApJ...787...84T}. These flares are short and powerful. Indeed, the observations of Crab pulsar nebula show flare duration between a few hours and up to several days with a rising/falling time scale of a few hours/days \citep{2012ApJ...749...26B,2013ApJ...765...52S}. This duration appears to be too large to be associated with the rotation of the neutron star (period of $33$ ms), and on the other hand too short for typical nebula dynamical time scale ($\approx 1$ year).

Relativistic magnetic reconnection is usually believed to be an efficient mechanism to explain these flares \citep{2012MNRAS.426.1374C}, possibly in a way similar to solar flares where magnetic energy is suddenly released and converted into other forms of energy. Indeed, the highly magnetized wind in the nebula entails a magnetic energy reservoir that is large enough to explain the typical observed flare energy ($\approx 10^{34} $J). However, achieving a fast enough time scale for releasing the magnetic energy is more tricky. For example, the standard view of  tearing instabilities developing in a single current sheet leads to a magnetic reconnection (with growth of magnetic islands) on a relatively slow time scale that is also very dependent on the (unknown) resistive dissipation \citep{2003ApJ...589..893L,2007MNRAS.374..415K}.
Recently, \citet{2011ApJ...737L..40U} and \citet{2012ApJ...746..148C} discussed that photons emitted by an extremely accelerated particle 
in an X-line may be able to explain the Crab flares, 
and \citet{2014ApJ...782..104C} calculated the evolution of a relativistic current sheet in the pulsar wind nebula using 3-dimensional PIC simulations. 
However, they did not succeed in reproducing the observed synchrotron spectra, in 
particular the 2011 April flare event. 

In \citep[Paper 1]{2013MNRAS.436L..20B}, an explosive mechanism is described, that is based on a magnetic reconnection process associated to the Double Tearing Mode (DTM). The favored site for the emission corresponds to a region situated in the stripped wind close to the light cylinder radius ($r \approx 50 r_L$). Indeed, the DTM is a Magnetohydrodynamic (MHD) instability that is well known to develop in multiple current sheets, as expected from the magnetic structure of pulsars magnetosphere with the presence of a current sheet wobbling around the equatorial plane \citep{1990ApJ...349..538C}. The explosive character of the mechanism is due to the sudden and fast development of a secondary instability that is structurally driven by the interaction of the initial magnetic islands situated on two successive current layers. It also manifests by a merging of the magnetic islands and is characterized by a bulk plasma flow reaching a magnitude close to the Alfv\'en speed.

More recently, the robustness of the mechanism has been shown by performing new Relativistic MHD simulations of the DTM of magnetically dominated plasmas, shifting to more realistic high magnetization parameter ($\sigma$) and magnetic Lundquist number ($S$ which is a measure of the dissipation) than previously explored in Paper 1 \citep[also referenced as Paper 2 below]{2015PPCF...57a4034P}. The results indicate a weak dependence of the time scale of the explosive phase on the magnetic Lundquist number and the magnetization parameter, $S^{0.2-0.3}$ and $\sigma^{0.3}$ respectively.

In this paper, with the aim to improve our model, we extend the previous results obtained in the two-dimensional approximation, by adding the effect of a magnetic guide field component perpendicular to the main plane. Such structure is indeed expected from a pulsar current sheet structure when the rotation and magnetic axis do not coincide. Indeed, the magnetic field structure in the striped wind can be described by an analytical expression given by \cite{2013MNRAS.434.2636P}. The main magnetic component is toroidal, along the azimuthal direction~$B_\varphi$ whereas the poloidal part is directed along the radius~$B_r$. Because of this particular magnetic configuration, we get $B_\varphi\propto \rlight/r$ and $B_r\propto (\rlight/r)^2$ both having the same magnitude at the light cylinder. Therefore the effect of a guide field in the pulsar striped wind should become negligible for distances much larger than~$\rlight$. We re-examine the scaling laws dependence with $\sigma$ and $S$, with a special emphasis on the effect of plasmoid-chain, that can develop in a transient way in the high $S$ regime. We also investigate the energetics of the reconnection process on the basis of our approximate magnetized fluid model. More precisely, the change of the temperature distribution during the explosive phase is determined in order to deduce an effective thermal particles spectrum, and the ensuing synchrotron emission. The results are then compared to the observations.


The paper is organized as follows. The numerical set-up is described in Section \ref{sec:setup}. The results of the DTM simulations are presented in Section \ref{sec:results}. Section \ref{sec:radiation} is devoted to the computation of the synchrotron spectra deduced from the temperature profile during the explosive reconnection event. Before ending with the conclusions in Section \ref{sec:conclusions}, we discuss the implications of our work for the relevant parameters in the striped wind, (sheet thickness and other parameters) Section~\ref{sec:discussion}.

\section{Numerical Setup}
\label{sec:setup}

\begin{table}
 \centering
  \caption{List of the parameters.}
  \begin{tabular}{lcccccc}
  \hline
   Name & $\sigma$ & $S$ & $L_y$ & $N_y$ & $B_{\rm G}/B_0$ & $c_{A,x}/c$ \\
 \hline
     SA1 & 0.2 & 200 & 10 $l$ & 1024 & 0 & 0.408 \\
     SA2 & 0.4 &  &  &  & & 0.535 \\
     SA3 & 0.6 &  &  &  & & 0.612 \\
     SA4 & 1.2 &  &  &  & & 0.739 \\
     SA5 & 12 &  &  &  & & 0.961 \\
     SA6 & 120 &  &  &  & & 0.996 \\
     SB1 & 0.2 & 3200 & 10$l$ & 2048 & 0 & 0.408 \\
     SB2 & 0.4 &  & 10$l$ & 2048 & & 0.535 \\
     SB3 & 0.6 &  & 20$l$ & 4096 & & 0.612 \\
     SB4 & 1.2 &  & 20$l$ & 4096 & & 0.739 \\
     SB5 & 12 &  & 20$l$ & 4096 & & 0.961 \\
     SB6 & 120 &  & 20$l$ & 4096 & & 0.996 \\
     RA1 & 0.4 & 400 & 10$l$ & 1024 & 0 & 0.535 \\
     RA2 &  & 800 &  & 1024 & & \\
     RA3 &  & 1600 &  & 1024 & & \\
     RA4 &  & 3200 &  & 2048 & & \\
     RC1 & 12 & 400 & 10 $l$ & 1024 & 0 & 0.961 \\
     RC2 &  & 800 & & 1024 & & \\
     RC3 &  & 1600 &  & 1024 & & \\
     RC4 &  & 3200 &  & 2048 & & \\
     GA1 & 12 & 1600 & 20$l$ & 2048 & 0.01 & 0.961 \\
     GA2 &  &  &  &  & 0.1 & 0.956\\
     GA3 &  &  &  &  & 0.5 & 0.859\\
     GA4 &  &  &  &  & 1 & 0.679 \\
     GA5 &  &  &  &  & 1.5 & 0.533  \\
     GB1 & 0.2 & 1600 & 20$l$ & 2048 & 0.5 & 0.365 \\
     GB2 & 1.2 &  &  &  &  & 0.661\\
     GB3 & 12  &  &  &  &  & 0.859\\
     GB4 & 120 &  &  &  & & 0.891 \\
     GC1 & 12 & 200 & 20$l$ & 2048 & & 0.859 \\
     GC2 &    & 800 &  &  & & \\
\hline
\end{tabular}
\end{table}

The evolution of the 2-dimensional (2D) double current sheets is modeled by using the relativistic resistive magnetohydrodynamic (RRMHD) approximation. The RRMHD equations are solved  by using  a numerical method developed by \citet{2011ApJ...735..113T} which solves  the equations in a conservative form using a finite volume method. 
Magnetic field is updated with the constrained transport algorithm \citep{1988ApJ...332..659E} 
which allows us to treat the divergence free magnetic field. 
In this paper, the numerical set up follows the Paper 1~\citep{2013MNRAS.436L..20B}, with an extra
magnetic field component $B_{\rm G}$ added to model the guide field. 

The initial current sheets are described by assuming the static relativistic Harris current sheets. 
The initial magnetic field profile is
\begin{equation}
  {\bf B} = B_0 \left[1 + \tanh [(y - y_0)/ l] - \tanh [(y + y_0)/ l] \right] {\bf x} + B_{\rm G} {\bf z}
  , 
  \label{eq:2.1}
\end{equation}
and the initial density profile is 
\begin{equation}
  \rho = \rho_{\rm b} + \rho_0 \left[1 / \cosh^2 [(y - y_0) / l] + 1 / \cosh^2 [(y - y_0) / l] \right]
  , 
  \label{eq:2.2}
\end{equation}
where $y_0 = 3 l$ is the half separation of the current sheets, and $l$ is the half thickness of the sheets. The gas pressure in the sheets is determined by the pressure balance. Note also that, for the sake of simplification, the guide magnetic field component is assumed to be uniform. The initial temperature $T$ is assumed to be uniform and constant, $k_{\rm B} T / m c^2 = 1$ where $k_{\rm B}, m, c$ are the Boltzmann constant, particle mass, and light velocity, respectively. 
The corresponding sound velocity is $c_s \simeq 0.516 c$. 
The upstream background density $\rho_{\rm b}$ is determined by the magnetization parameter $\sigma \equiv B^2/4 \pi \rho h c^2$ where $h$ is the specific enthalpy. The relativistic ideal equation of state is assumed, in normalized units $h = 1 + \Gamma/(\Gamma-1)p/\rho c^2$ where $\Gamma = 4/3$. The resistivity coefficient $\eta$ is assumed to be uniform and constant, and is related to the Lundquist number $S$ through $S = 4 \pi c l / \eta$ ($4 \pi$ is necessary due to the Gauss unit.). Note that in this paper, we use the light speed as the characteristic velocity to define our Lundquist number.

The simulation domain is bounded by a rectangular domain
of dimensions  $[0, L_x] \times [-L_y, L_y]$, with a fixed size value $L_x = 20 l$.
As the boundary condition of $x$-direction is taken to be periodic, the $L_x$ value is precisely chosen
in order to get a linearly unstable mode with a wavelength close to
to the fastest one according to MHD stability analysis (see Papers1-2 for more detail). 
The $L_y $ value is taken to be large enough to prevent the boundary effects on the sheets evolution, as
free boundary conditions are used in this lateral $y$-direction.
We divide this simulation domain into homogeneous numerical square meshes with size $L_y / N_y$. 

A small divergence-free magnetic field perturbation is also added (at $t=0$) as a vector potential $\delta A_z$ at the cell-edges
in order to trigger the system evolution: 
\begin{equation}
  \delta A_z = - B_1 \frac{L_x}{2 \pi} \cos \left[\frac{2 \pi}{L_x} (x - L_x/2) \right] \exp [- (|y| - y_0)^2]
  ,
  \label{eq:2.3}
\end{equation}
where $B_1 = 3 \times 10^{-4} B_0$ is the perturbation amplitude.

The parameters $\sigma, S, L_y, N_y, B_{G}$ used in the simulations are given in Table 1 
\footnote{
Note that in this paper, we treated high-Lundquist number plasmas. 
In general, it is very difficult to find the numerical convergence in the presence of plasmoid-chain due to its turbulent nature. 
In this paper, we have checked the convergence of the maximum velocity and temperature at the explosive phase of DTM, 
which is given in Appendix A. 
}.
The actual numerical value of the parameters, such as $\rho_0, \rho_b, B_0$, follow Paper I. 

\section{Results}
\label{sec:results}

As previously reported in Paper 1, the evolution of the system exhibits
four distinct stages: (1) an initial linear growth of two magnetic islands 
arranged asymmetrically in the two current layers and developing on
a time scale $S^{1/2}$; (2) an ensuing saturation on the slow diffusion
time scale $S^1$; (3) an explosive growth of the islands interacting with a triangular
deformation and driving a fast magnetic reconnection event characterized by a merging between the
two islands; (4) a relaxation towards a new stable state. 

This overall behavior does not change in highly magnetized plasma regime with large Lundquist numbers,
as reported in Paper 2 \citep{2015PPCF...57a4034P}.
An important result for the robustness of our model is that, a relatively weak dependence of the time scale of the explosive phase
on the magnetic Lundquist number ($S^{0.2-0.3}$) and on the magnetization parameter ($\sigma^{0.3}$) is obtained.

Interestingly, the explosive stage is
however characterized by a transient growth of smaller magnetic islands identified as plasmoid-chain instability. 
The plasmoids appear when a critical local Lundquist number is reached,
as it is the case for runs using $S = 3.2 \times 10^3$ (see Fig. 1 in Paper 2). Thus, in order to quantify the effect
of the plasmoid-chain, we re-examine in somewhat more details the scaling law dependence of the explosive
stage.

\subsection{Scaling law dependence and plasmoid chain effect}
\setcounter{figure}{0}

Following the Papers 1-2, we use the maximum 4-velocity in $x$-direction, $u_x = \gamma V_x$, as a characteristic variable of the double tearing mode
evolution.

\begin{figure}
 \centering
  \includegraphics[width=8.cm,clip]{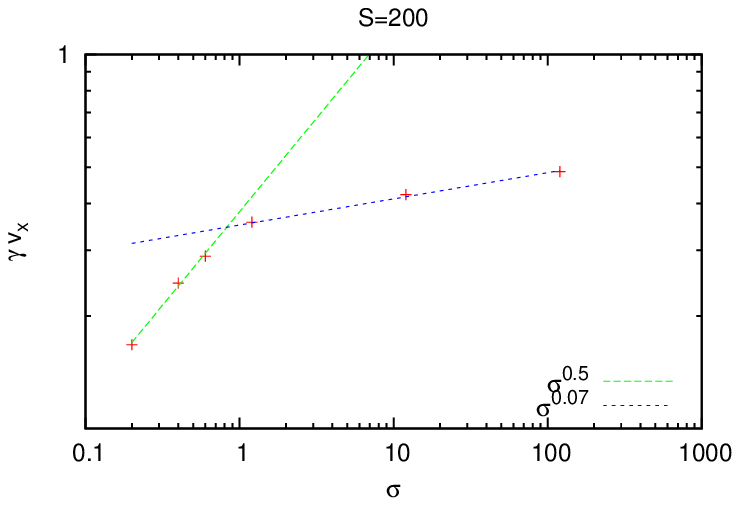}
  \includegraphics[width=8.cm,clip]{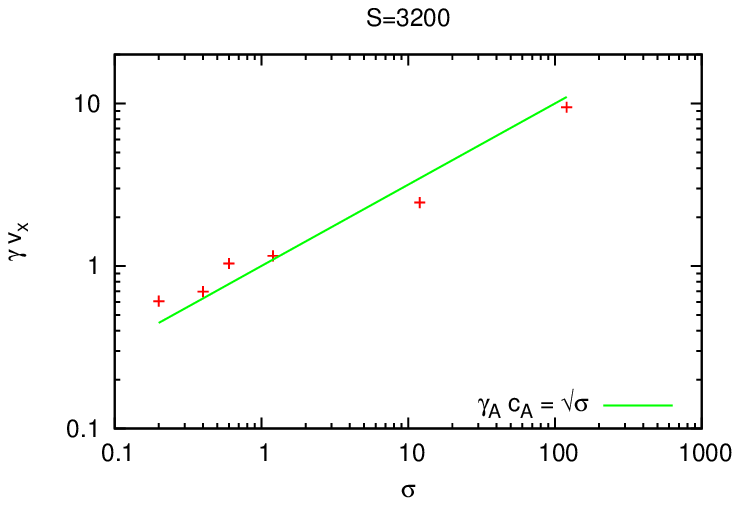}
  \caption{Maximum 4-velocity $\gamma V_x$ of double tearing mode with respect to the magnetization parameter $\sigma$. Top: Low Lundquist number cases, $S = 200$: 
runs SA1-6. Bottom: Large Lundquist number cases, $S=3200$: runs SB1-6. 
    }
  \label{fig:3.2.1}
\end{figure}

Firstly, we examine the scaling law of $u_x$ on the magnetization parameter for different Lundquist number values. The results are shown in Panels of Figure \ref{fig:3.2.1}. The top and bottom panels are for the low and large Lundquist number cases, respectively. In the low Lundquist number case, 
runs SA1-6, we found that the maximum 4-velocity scales as $\sim \sigma^{0.5}$ and  $\sim \sigma^{0.07}$ in the small $\sigma$ and large $\sigma$ regimes, respectively\footnote{
Note that the obtained scaling law on $\sigma$-parameter is a little different from our previous work which indicated $\sim \sigma^{0.3}$. 
This is mainly due to including $\sigma=0.2$ which is actually included in a new scaling regime we found for the first time in this paper. 
}. This indicates that the maximum 4-velocity in the low-$\sigma$ region is proportional to the Alfv\'en 4-velocity: $\gamma_A c_A = \sqrt{\sigma}$.  However, we note that the obtained maximum velocity is not the Alfv\'en velocity but approximately just half of it. This is consistent with the outflow velocity of the single relativistic tearing instability with small Lundquist number, indicating the deceleration by the large enthalpy in the outflow region~\citep{2011ApJ...739L..53T}. The bottom panel of Figure 1 shows the maximum 4-velocity $\gamma V_x$ in the case of $S=3200$, 
runs SB1-6. Differently from the low Lundquist number case, the maximum 4-velocity can be well reproduced by the Alfv\'en 4-velocity, $\sqrt{\sigma}$. This is because, in the large Lundquist number case, the sheets becomes plasmoid-chain as reported in Paper 2, 
which includes the Petschek type structure \citep{2012PhPl...19i2110B,2013ApJ...775...50T}.

\begin{figure}
 \centering
  \includegraphics[width=8.cm,clip]{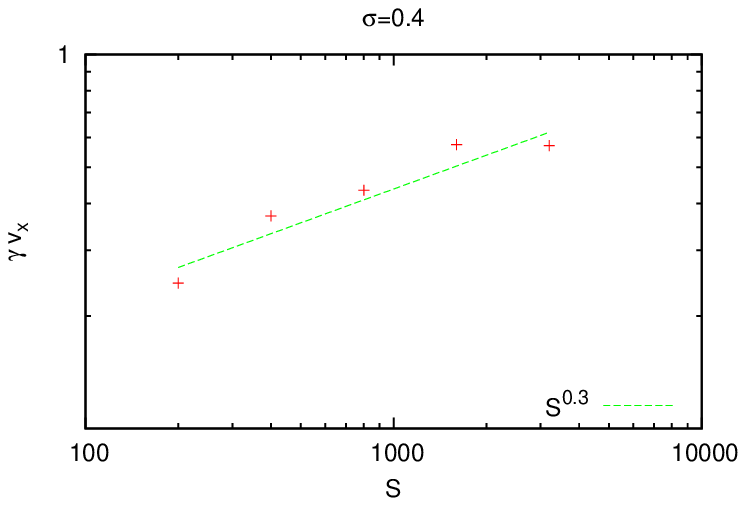}
  \includegraphics[width=8.cm,clip]{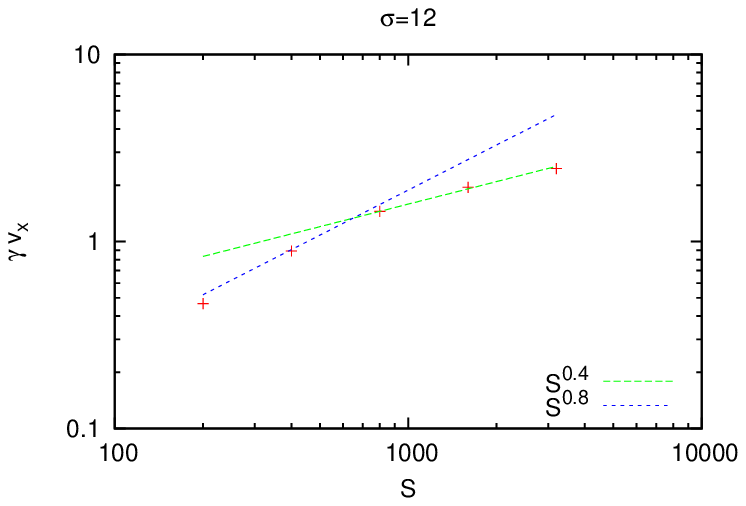}
  \caption{Maximum 4-velocity $\gamma\,V_x$ of double tearing mode with respect 
runs RA1-4, SA2. Bottom: High magnetization cases, $\sigma = 12$: runs RC1-4, SB5.} 
  \label{fig:3.2.2}
\end{figure}

Next, we examine the scaling law of $\gamma V_x$ on the Lundquist number. The numerical results are reported
in Panels of Figure \ref{fig:3.2.2}. The top and bottom panels are in the small and high $\sigma$ cases, respectively. 
In the small $\sigma$ case, the maximum velocity scales approximately as $S^{0.3}$ as reported in the Paper 1. 
In the large $\sigma$ case, the maximum 4-velocity scales as $S^{0.3-0.4}$ as in the small Lundquist number regime. 
However, the scaling becomes $\sim S^{0.8}$ and gradually saturates in the large Lundquist number regime. 
This is because the maximum velocity becomes very close to the Alfv\'en 4-velocity in the upstream region, $\sqrt{\sigma} \simeq 3.5$, when $\sigma=12$.

\begin{figure}
 \centering
  \includegraphics[width=8.cm,clip]{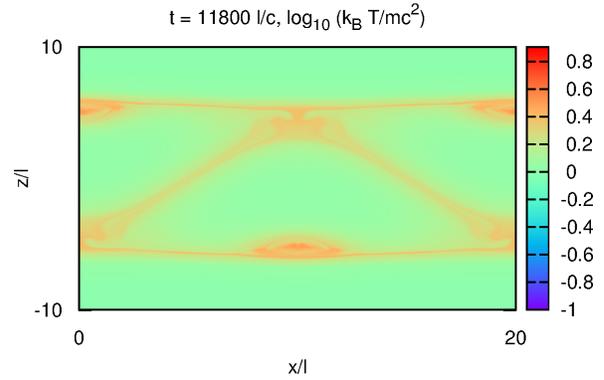}
  \caption{
           A snapshot of temperature profile at the burst phase of run GA4. 
           }
  \label{fig:3.2.4.1}
\end{figure}

\begin{figure}
 \centering
  \includegraphics[width=8.cm,clip]{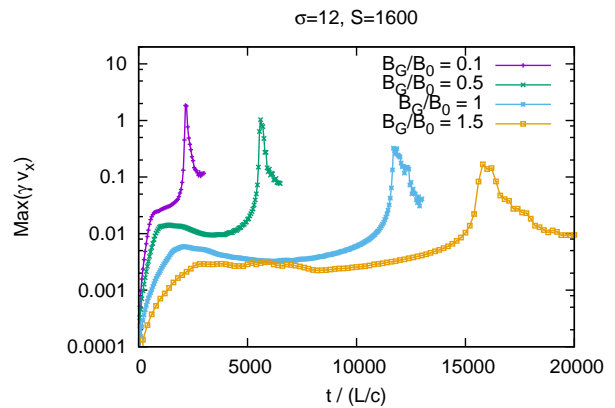}
  \caption{
           Temporal evolution of the maximum 4-velocity with different guide field. 
           }
  \label{fig:3.2.4.2}
\end{figure}

\subsection{Guide magnetic field effects}

In order to investigate the effect of a magnetic guide field component, the direction of the total magnetic field is varied with a fixed magnitude, 
that is, ${\bf B} = B \cos \Phi {\bf x} + B \sin \Phi {\bf z} \equiv B_0 {\bf x} + B_G {\bf z}$ 
where $B$ is the constant total magnetic field strength, and $\Phi$ is the angle between x-axis and the magnetic field direction. 
The corresponding runs are GA1-GA5 in Table 1. 

Figure \ref{fig:3.2.4.1} is a snapshot of temperature profile of run GA4. 
The basic spatial structure is roughly similar to the case without guide field. 
However, the resulting temperature from the burst phase is much smaller than the case without guide field (see Figure \ref{fig:4.1.2}). 
This is because the compression of the guide field reduces the increase of gas pressure. 
Note also that the sheet region does not have plasmoid-chain, differently from the case without guide field as reported in Paper II. 
This is due to the stabilization of tearing instability by guide field \citep{1993SSRv...65..253S}, 
which prohibits the secondary-tearing instability leading to the plasmoid-chain. 
Figure \ref{fig:3.2.4.2} is the temporal evolution of the maximum 4-velocity in x-direction. 
It shows that the initial single tearing instability is slowed down by guide field as is predicted by \citet{1993SSRv...65..253S} in non-relativistic work. 
It also shows the evolution of secondary phase is slowed down and the maximum velocity becomes slower as increasing the guide field component. 
\begin{figure}
 \centering
  \includegraphics[width=8.cm,clip]{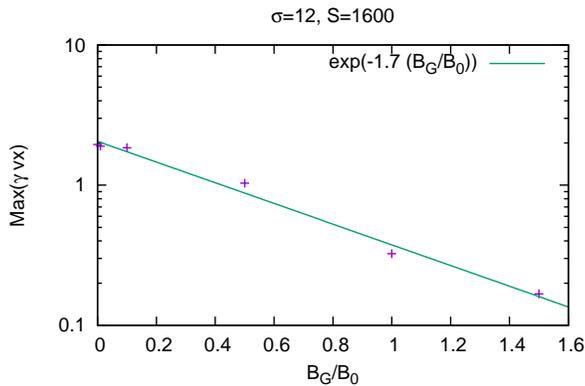}
  \caption{
    The maximum 4-velocity $\gamma\,V_x$: runs GA1-5. 
 }
  \label{fig:3.2.3.1}
\end{figure}

\begin{figure}
 \centering
  \includegraphics[width=8.cm,clip]{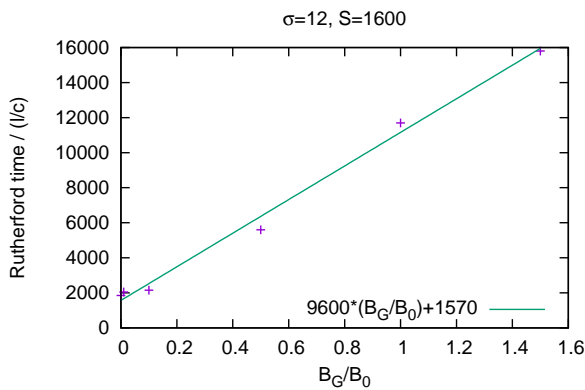}
  \caption{
    The Rutherford time (see text for the definition). The horizontal axes are the ratio of the guide field to the reconnecting magnetic field component.             
 }
  \label{fig:3.2.3.2}
\end{figure}

Figure \ref{fig:3.2.3.1} plots the guide field dependence of the maximum 4-velocity. 
In the figure, 
the maximum 4-velocity exhibits a decrease with increasing guide field as reported by \citet{2009ApJ...705..907Z} and \citet{2011MNRAS.418.1004Z}. 
This can be interpreted by a decrease of the reconnection magnetic field component $B_0$, which results in the decrease of the Alfv\'en velocity, $c_A = \sqrt{\sigma_x/(1 + \sigma)}$, along the sheets\footnote{Here, $\sigma_x \equiv B_x^2/4 \pi \rho h c^2 \gamma^2$ is the magnetization parameter in terms of the reconnection magnetic field component.}. This is due to the decrease of the reconnection magnetic field strength as: $B_0 = B \cos \Phi$. 
Figure \ref{fig:3.2.3.2}
shows the Rutherford time (see Papers 1-2), defined as the time at which the explosive phase is triggered. The results show that the Rutherford time increases linearly. Note that the slope is very steep, and the Rutherford time becomes even twice when $B_{\rm G} = 0.5 B_0$. 
Note that these properties are very similar to the behavior of the relativistic Petschek slow shocks with guide field~\citep{2005MNRAS.358..113L}. 

\begin{figure}
 \centering
  \includegraphics[width=8.cm,clip]{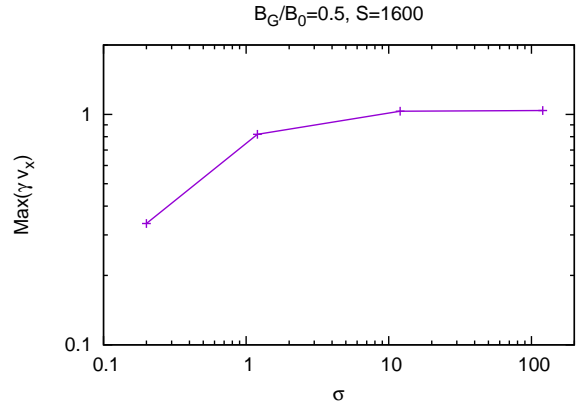}
  \includegraphics[width=8.cm,clip]{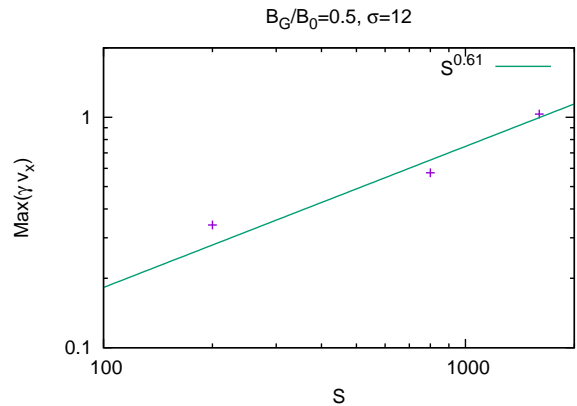}
  \caption{Top: Maximum 4-velocity dependence on $\sigma$, run GB1-4. Bottom: Maximum 4-velocity dependence on $S$, GC1,2,GB3.
           In these calculations, the ratio $B_G/B_0$ is fixed.}
  \label{fig:3.2.4}
\end{figure}

Figure \ref{fig:3.2.4} are the maximum 4-velocity dependence including guide field which is fixed as $B_G/B_0 = 0.5$. 
The top panel shows the magnetization parameter dependence. 
It shows the dependence is completely different from the no guide field case in Figure \ref{fig:3.2.1}, 
and the guide field drastically reduce the maximum 4-velocity, although the maximum 4-velocity slightly increases with $\sigma$. 
This is because the guide field hinders the evolution of plasmoid-chain as indicated in Figure \ref{fig:3.2.4}, 
and the sheets becomes usual relativistic Sweet-Parker sheet whose outflow saturates an upper limit before reaching the Alfv\'en velocity
~\citep{2011ApJ...739L..53T}. 
The bottom panel shows the Lundquist number dependence. 
It shows the maximum 4-velocity increases with Lundquist number similarly to no-guide field case shown in Figure \ref{fig:3.2.2}. 
However, 
the power-law index of the Lundquist number is around $0.6$ 
which is an intermediate value indicated in the bottom panel of Figure \ref{fig:3.2.2}, that is, between $0.4$ and $0.8$. 
These panels clearly show that 
the guide field basically weakens the energy conversion resulted from DTM, 
and the burst phase becomes less explosive even in highly magnetized plasma with high Lundquist number. 

\begin{figure}
 \centering
  \includegraphics[width=7.cm,clip]{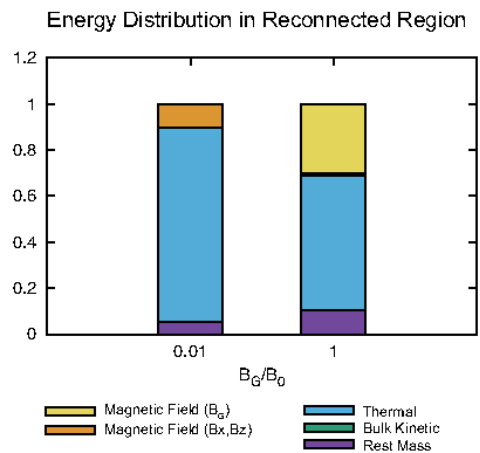}
  \caption{Energy density in the thermalized region after DTM burst with guide field component, corresponding to runs: GA2 and GA4.}
  \label{fig:3.2.5}
\end{figure}

Figure \ref{fig:3.2.5} is the energy distribution to each component in the reconnected region of runs GA2 and GA4. 
It shows most magnetic field energy is converted into the thermal energy in the both cases. 
However, the guide field energy occupy around 30\% of the total energy in GA4. 
This indicates the guide field magnetic field pressure compensates the compression by DTM, 
and reduces the conversion into thermal energy. 
This is basically consistent with the two-fluid result by \citet{2009ApJ...705..907Z}. 
However, the thermal energy is still dominant in the case of GA4, 
and this reflects the fact that DTM is not related to slow shocks. 
\section{DTM thermal energy spectrum and Crab Flares}
\label{sec:radiation}

In our previous works \citep{2013MNRAS.436L..20B,2015PPCF...57a4034P}, we did not discuss in detail the radiation properties associated to the double tearing mode evolution. In order to obtain more quantitative results about possible radiative signatures of the reconnection phase, in this section, we compute the thermal synchrotron spectrum expected to emanate from the relativistic and highly magnetized plasma during its explosive phase. The procedure is explained in the following lines, our aim being to compare our synthetic spectra with the variability of the Crab gamma-ray flares.

\subsection{Temperature Profile}

As already discussed in Paper 1,2, the double tearing mode shows an explosive reconnection phase during which temperature drastically increases in the reconnected region even in the case of moderate value of the magnetization parameter~$\sigma$. In this paper, we explore ultra-relativistic Poynting dominated flows assuming a high-Lundquist number for the plasmas, and show the resulting explosive reconnection becomes more violent, 
that is, the flow speed becomes faster and the temperature becomes higher, 
as the magnetization $\sigma$-parameter and the Lundquist number are increased. Since the Crab pulsar wind is considered to be a high-$\sigma$ plasma with very large Lundquist number, we expect the double tearing mode to be sufficiently energetic to explain the flares in the Crab pulsar/nebula system.

\begin{figure}
 \centering
  \includegraphics[width=8.cm,clip]{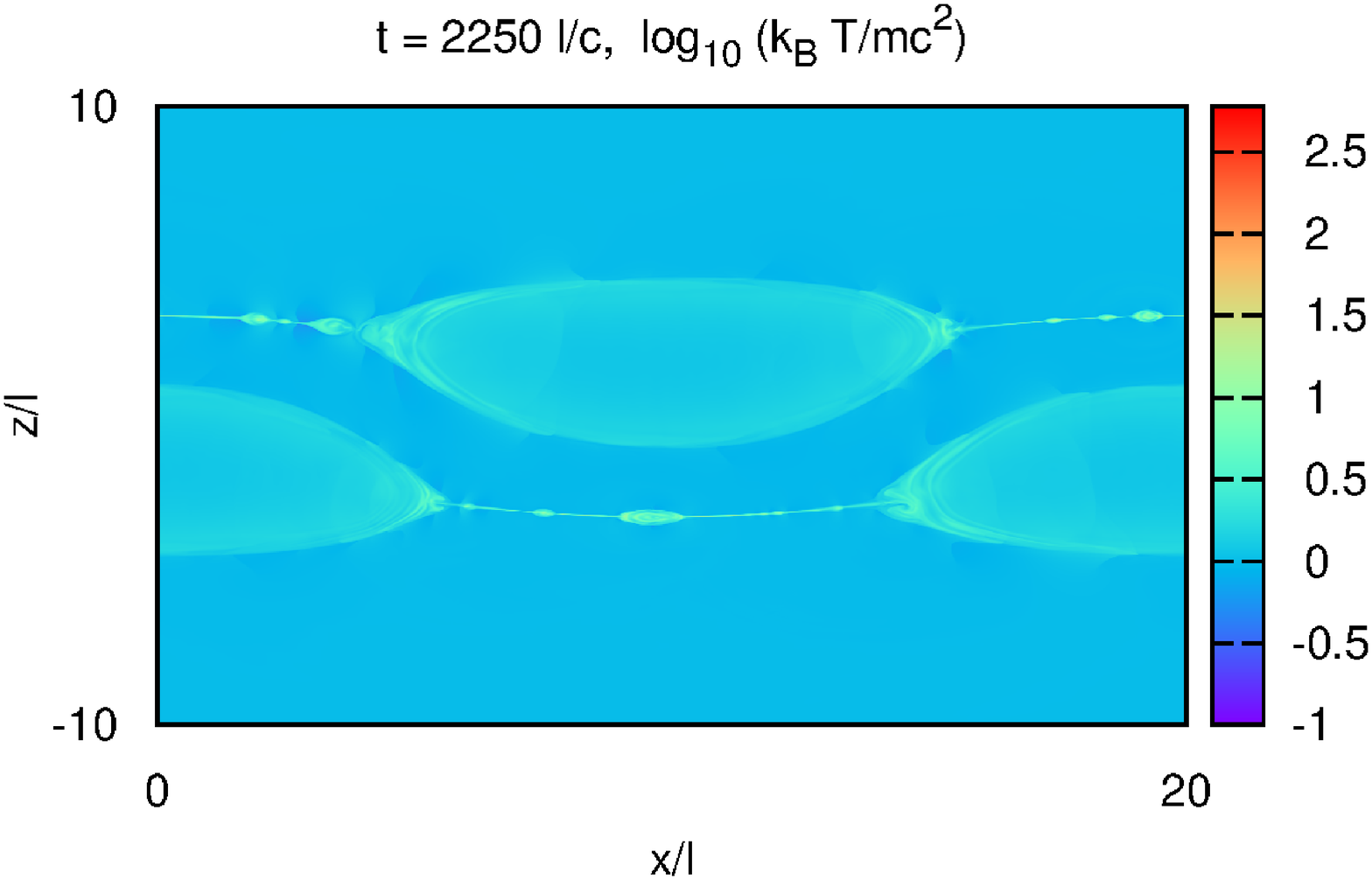}
  \includegraphics[width=8.cm,clip]{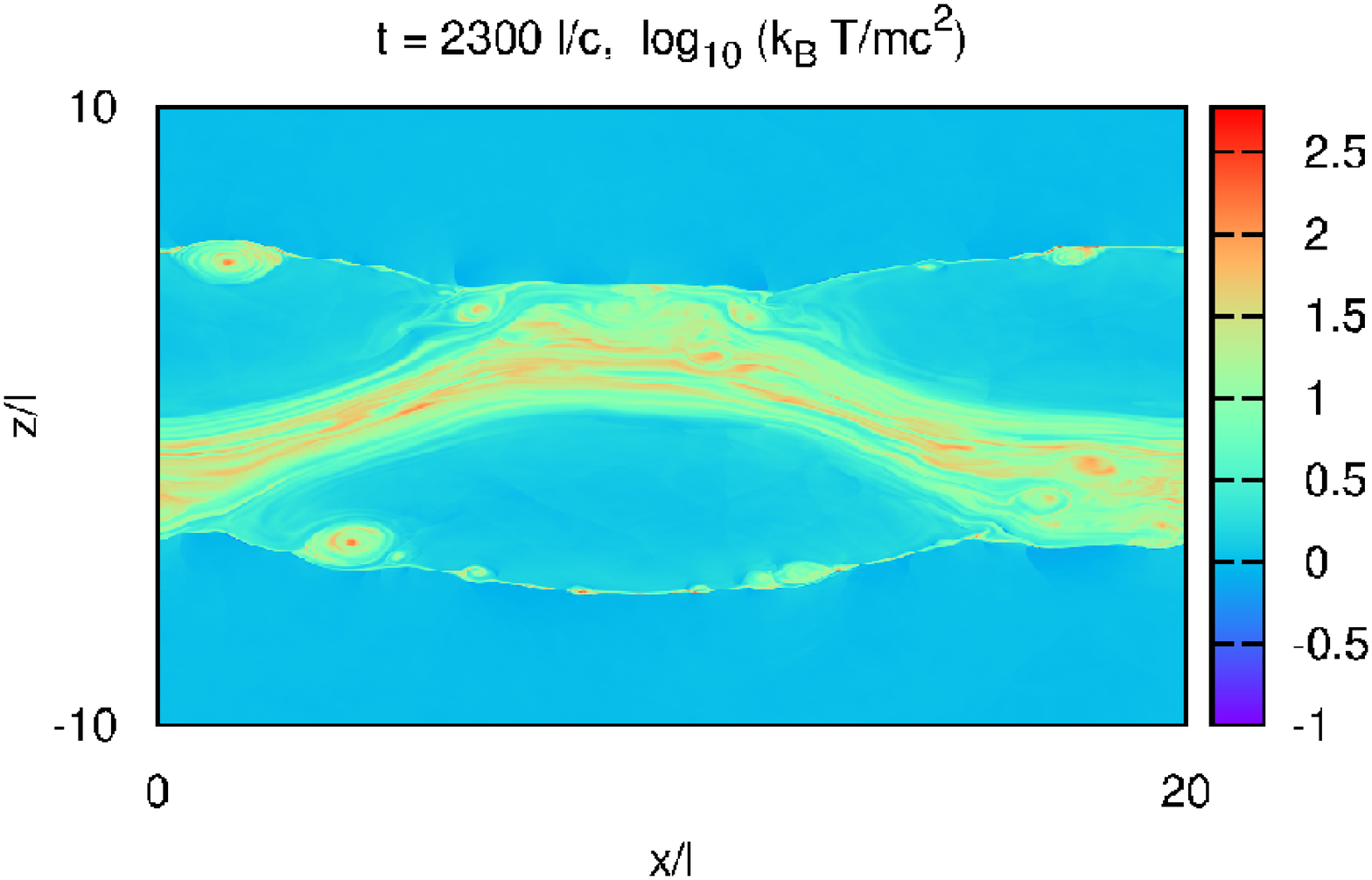}
  \caption{The temperature profiles $\log_{10} [k_{\rm B} T / m c^2]$ during the double tearing burst phase of the run SB6. The top and bottom panel show just before and after the burst phase, respectively. }
  \label{fig:4.1.1}
\end{figure}

Figure \ref{fig:4.1.1} shows the profile of temperature just before and after the explosive phase of double tearing mode in a highly magnetized plasma, $\sigma=120$, with high Lundquist number, $S=3200$. As is indicated in 
Figure \ref{fig:4.1.1}, 
the high Lundquist number plasma gives very narrow reconnection sheets in which several plasmoids can be observed (see also our Paper 2). The plasmoid region becomes very dense and also very hot with temperature as high as $k_B T \gtrsim \sigma mc^2$ because of the compression by nearly Alfv\'enic reconnection flows along the sheet~\citep{2013ApJ...775...50T}. After the explosive phase, the magnetic field between the two original sheets is forced to reconnect, and dissipate its energy into kinetic bulk flow and thermal energy. 
Since the motion inside of the reconnected region becomes highly stochastic, the resulting kinetic bulk flow energy rapidly dissipated into the thermal energy. In particular, the random motion induces strong compression, and this also increase the temperature in the plasmoids.

\subsection{Synchrotron Energy Spectrum}

\begin{figure}
 \centering
  \includegraphics[width=8.cm,clip]{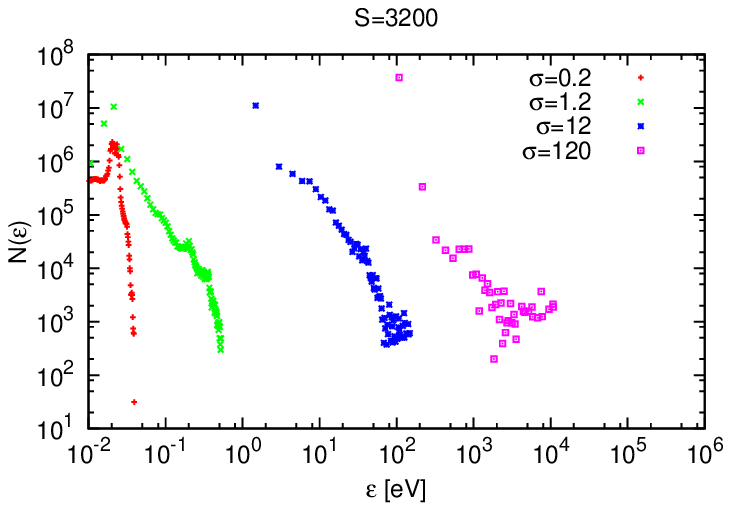}
  \includegraphics[width=8.cm,clip]{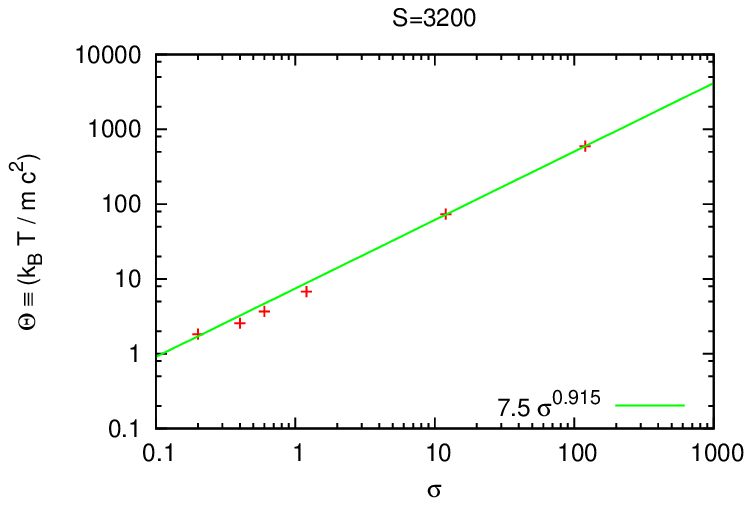}
  \caption{Top: The energy spectrum with various magnetization parameter $\sigma$. Bottom: maximum temperature with respect to background magnetization parameter. In both case, $S=3200$ is assumed, and the pulsar parameters are $r=50r_{L}$ and $\Gamma_W = 300$. The runs are corresponded to runs SB1-6, 
          and the data at the burst phase reaching maximum 4-velocity is used. 
  }
  \label{fig:4.1.2}
\end{figure}

Next, we investigate the synchrotron energy spectrum using our numerical results. 
The typical photon energy for the synchrotron emission can be written as: 
\begin{equation}
  \label{eq:4.1}
  \epsilon_{\rm sync} = \frac{3}{2} \gamma^2 \frac{B}{B_q} m c^2
  ,
\end{equation}
where $B$ is the magnetic field strength as measured in the frame where $\epsilon_{\rm sync}$ is detected, $B_q = m^2 c^3/e \hbar$ is the critical magnetic field, $m$ the electron mass, $e$ its electric charge, $\gamma$ the typical Lorentz factor of the electrons and $\hbar$ the reduced Planck constant. Assuming the synchrotron radiation is mainly emitted by the pair plasma which is in local thermal equilibrium due to the MHD approximation, the Lorentz factor in Equation (\ref{eq:4.1}) can be expressed for an ultra-relativistic plasma equation of state as: $\gamma \sim 3 k_{\rm B} T / m c^2 \equiv 3 \Theta$. 
For later convenience, we introduced the normalized temperature $\Theta$. Note that the temperature~$T$ depends on the location in the simulation box. In the explosive phase, a strong gradient are formed leading to drastic variation in the temperature profile from point to point as seen in Figure~\ref{fig:4.1.1}. 
In the pulsar striped wind region, the dominant background magnetic field component is toroidal and propagates as an entropy wave thus decreasing with radius~$r$ according to, $B_0 = B_{\rm L} \, \rlight/r$, 
where $\rlight \sim 1.5 \times 10^6$ [m] is the radius of the light cylinder and $B_{\rm L} \sim 100$[T] the magnetic field strength at $\rlight$ for the Crab pulsar.
The synchrotron photon energy in the fluid comoving frame becomes
\begin{align}
  \label{eq:4.4}
  \bar{\epsilon}_{\rm sync} &\sim 1.73 \times 10^{-4} [{\rm eV}] \left(\frac{3 k_{\rm B} T}{m c^2}\right)^2 \left( \frac{\bar{B}}{1[{\rm T}]} \right)
  \nonumber                           
  \\
  &= 0.156~[{\rm eV}] \Theta^2 \Gamma_{\rm W}^{-1} \left(\frac{r}{\rlight}\right)^{-1}
  \left( \frac{\bar{B}}{\bar{B}_0} \right) \left( \frac{B_L}{100[{\rm T}]} \right)
  ,
\end{align}
where $\bar{B}_0 = B_0 / \Gamma_{\rm W}$ is the background magnetic field measured in the fluid comoving frame, or the simulation frame and $\Gamma_{\rm W}$ is the Lorentz factor of the pulsar wind. Finally, the Lorentz transformation into the pulsar rest frame (observer frame) gives us 
\begin{equation}
   \label{eq:4.4a}
  \epsilon_{\rm sync,L} = \frac{\bar{\epsilon}_{\rm sync}}{\Gamma_{\rm W}\,(1-\beta_{\rm W}\,\cos\vartheta)} \equiv \delta_{\rm W}~\bar{\epsilon}_{\rm sync}
\end{equation}
where $\vartheta$ is the angle between the velocity of the reconnecting blob and the observer line of sight, 
$\beta_{\rm W}$ is the velocity of the wind in the unit of light velocity, 
and $\delta_{\rm W}$ is the ``\textit{Doppler factor}''. 
If we assume that this blob is pointing towards Earth in the most favorable case, $\delta_{\rm W} \simeq 2 \Gamma_{\rm W}$, 
the peak energy of the photons will be
\begin{equation}
  \label{eq:4.5}
  \epsilon_{\rm sync,L} \sim 0.311 [{\rm eV}]~ \Theta^2 \left( \frac{r}{\rlight} \right)^{-1}
  \left( \frac{\bar{B}}{\bar{B}_0} \right) \left( \frac{B_L}{100[{\rm T}]} \right)
\end{equation}
The energy spectrum can be obtained by integrating Equation (\ref{eq:4.5}) over the numerical spatial domain, $\int dV n \epsilon_{\rm sync}$, where $n$ is the number density of emitting particles 
in the laboratory frame
. 
The top panel of Figure \ref{fig:4.1.2} shows the energy spectra with several value for the magnetization parameter $\sigma$ 
corresponding to the Runs SB1-6 at the burst phase when reaching maximum 4-velocity
. Typical values for the Crab pulsar are: $r = 50 \rlight$~\footnote{
Here, the emission region is assumed to start approximately at a fixed radius outside the light-cylinder following the prescription given by \citep{2005ApJ...627L..37P}. 
This is a necessary requirement for both fitting the synchrotron spectrum to the observed value (as the synchrotron emissivity decreases with distance) 
and adjusting the timescale of the Crab flares taking into account time dilation. 
Although this makes it difficult to discuss timescale of 
DTM, 
we use the DTM timescale as the observed flare timescale in Equation (\ref{eq:4.7}) 
since this is one of the characteristic timescales of DTM model in the striped wind. 
The validity of the above assumption will be investigated in a future work. 
}
 and $\Gamma_{\rm W} = 300$~\footnote{
This is twice larger than the value inferred in Paper 2, $\Gamma_{\rm W} \lesssim 150$. 
This is due to changing of the values of some parameters in the dynamical time of the explosive phase, Equation (\ref{eq:4.7}), 
in order to explain the cut-off energy of the observed flare. 
}. 
This panel clearly shows the energy in their body part increases with $\sigma$-parameter, approximately 100 times increase as $\sigma$-parameter increases by one order of magnitude. This can be understood from the bottom panel of Figure \ref{fig:4.1.2}. The panel is the maximum temperature during the explosive phase in terms of the $\sigma$-parameter. It shows the maximum temperature increases linearly with the $\sigma$-parameter. Since the photon energy depends on the temperature as $T^2$, as indicated in Equation (\ref{eq:4.5}), the resulting energy spectrum can be estimated as $T^2 \propto \sigma^2$. The observed Crab flares have their mean energy around $10^2$ to $10^3$ [MeV] energy region~\citep{2012ApJ...749...26B,2014RPPh...77f6901B}, so that the double tearing mode can explain the Crab flares if the Crab pulsar wind has $\sigma \gtrsim 10^5$ which is in agreement with current estimates from theoretical expectations~\citep{2009ASSL..357..421K} about a trans-Alfv\'enic flow. We also note that the energy spectrum in Figure \ref{fig:4.1.2} increases with $\sigma$-parameter. This is because, in high $\sigma$ cases, the number of plasmoids along the sheets increases 
as reported in \citep{2013ApJ...775...50T}; 
In plasmoid region, the plasma is characterized by a high temperature and a high density, which increases the high energy photons and make the energy spectrum harder
\footnote{Our numerical results in the case of $\sigma=120$ gave us an effective index in the high energy region $\gamma_{\rm F,DT} \sim 1.3$, 
which is not so far away from the observed energy spectral index of the flaring component $\gamma_{\rm F,obs} = 1.27 \pm0.12$.
Comparing with the non-thermal PIC simulation results obtained by \citet{2014ApJ...782..104C}, 
the DTM spectrum is harder in high energy region due to the plasmoids.
}
.
Note that the plasma is far below the radiation reaction limit because in the wind frame the energy of the synchrotron photons is much less that the limit of 240~MeV. Therefore, the dynamics of the plasma presented in this paper is not significantly perturbed by the radiative losses. Our RRMHD code does not include any such losses so far. We computed the spectra by a post-processing algorithm. 

Finally, the synchrotron cooling time in the wind comoving frame can be written as: 
\begin{align}
  \bar{\tau}_{\rm sync} =& \frac{3}{4} \frac{m c}{\gamma \sigma_{\rm T} U_{\rm B}} 
  \nonumber
  \\
  \simeq& 2.3 \times 10^7~[{\rm s}]
  \nonumber
  \\
  &\times \left(\frac{\Gamma_{\rm W}}{300}\right)^2 \Theta^{-1} \left( \frac{\bar{B}}{0.05 \bar{B}_0} \right)^{-2} 
  \left( \frac{r}{50 r_{\rm L}} \right)^2 \left( \frac{\bar{B}_{\rm L}}{100 [{\rm T}]} \right)^{-2}
  \label{eq:4.6}
  ,
\end{align}
where $\sigma_{\rm T}$ is the Thomson scattering cross section and $\bar U_{\rm B}$ is the magnetic field energy density in the fluid comoving frame. 
Note that this timescale is estimated in the merged sheet where the magnetic field release occurs. 
In our simulations, the dynamical time of the explosive phase in the simulation frame, $\bar{\tau}_{\rm dyn}$, is about $50 l/c$. 
Using $l = 2 \pi \alpha \Gamma_{\rm W} r_{\rm L}$, 
\footnote{$\alpha$ is a ratio between the sheet width and the sheet separation 
(see also Figure \ref{fig:4.3.1}).} 
it reduces to 
\begin{equation}
  \bar{\tau}_{\rm dyn} = 25~[{\rm s}] \left(\frac{\alpha}{0.05} \right) \left(\frac{\Gamma_{\rm W}}{300}\right) \left( \frac{P}{33[{\rm ms}]} \right)
  \label{eq:4.7}
  ,
\end{equation}
where $P = 2 \pi r_{\rm L}/c$ is the rotation period of the central neutron star\footnote{
Note that if we consider the Lorentz-time dilation this gives a flare timescale: $\tau_{\rm dyn,lab} \sim 2$ hours in the pulsar rest frame, 
which is within the observed flare timescale, 8 hours. 
}. 

Comparing these 2 timescales, we obtain 
\begin{align}
  \frac{\bar{\tau}_{\rm sync}}{\bar{\tau}_{\rm dyn}} 
  = &9.4 \times 10^5 \left(\frac{\Gamma_{\rm W}}{300}\right) \Theta^{-1} \left( \frac{\bar{B}}{0.05 \bar{B}_0} \right)^{-2} 
  \left( \frac{r}{50 r_{\rm L}} \right)^2 
  \nonumber
  \\
  &\times \left( \frac{\bar{B}_{\rm L}}{100 [{\rm T}]} \right)^{-2} \left(\frac{\alpha}{0.05} \right)^{-1} \left( \frac{P}{33[{\rm ms}]} \right)^{-1}
  \label{eq:4.8}
  .
\end{align}
This indicates the maximum energy by DTM in the Crab pulsar wind would be obtained when $\Theta \sim 10^6$. 
Note that we obtain a cutoff energy scale around 300 MeV by substituting this temperature into Equation (\ref{eq:4.5}) assuming $\bar{B} \sim 0.05 \bar{B}_0$ and $r = 50 r_{\rm L}$, 
which roughly reproduces the observed cutoff energy, 350MeV. 
Figure \ref{fig:4.1.2} also indicates this temperature can be obtained when $\sigma \sim 10^5$. 


\section{Applicability to Crab Pulsar Wind}
\label{sec:discussion}


\begin{figure}
 \centering
  \includegraphics[width=6.cm,clip]{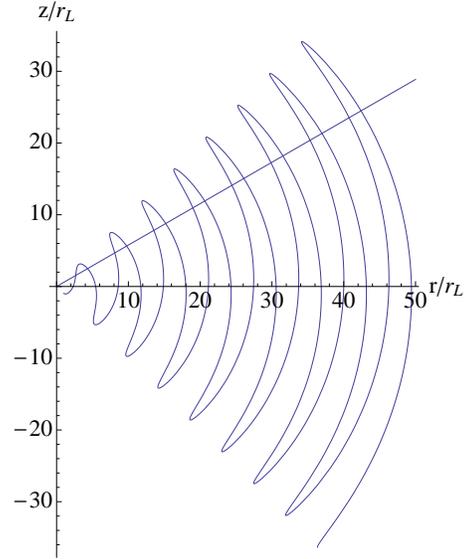}
  \caption{A meridional profile of current sheets in the striped wind. 
           The inclination angle is assumed $\pi/4$, 
           and the light cylinder radius $r_{\rm L}$ is used for the unit of each axis. 
           A solid line is $z = x \tan \theta$ where $\theta = \pi/6$ 
           along which the distance between the sheets becomes imbalanced. 
          }
  \label{fig:4.3.1}
\end{figure}

In this paper, we discuss the double tearing mode to explain the Crab GeV flares. For the simulation setup, we assume a separation of two consecutive current sheets to be around 3 times of their thickness. Although the sheet separation being fairly confidently constrained to be of the order of the light cylinder radius using the geometry of the striped wind model~\citep{1971CoASP...3...80M}, there is still no accepted theoretical estimate giving the  thickness of one sheet. However, from an observational point of view, the Fermi Large Area Telescope (Fermi/LAT) detected more than hundred gamma-ray pulsars showing small scale details in their light curves in the GeV band \citep{2010ApJ...708.1254A}. If we assume that the striped wind current sheets are responsible for the gamma ray pulses, the width of each pulse should reflect the thickness of the sheet itself. Indeed, for sufficiently relativistic outflows, the spiral structure combined to relativistic beaming effects gives rise to pulsed emission. If the current sheet has a half thickness of~$l$ then the half width of the corresponding pulse should be $l/\upi\,\beta\,\rlight$ where $\beta\approx1$. For the Crab pulsar, the pulse width is about 10\% of the pulsar period thus the underlying current sheet thickness should be around 10\% of the striped wind wavelength\citep{2005ApJ...627L..37P,2012MNRAS.424.2023P}. 
We consider this supports our MHD treatment of the double current sheets 
since the indicated sheet width is much larger than the kinetic scales, such as the electron's Larmor radius. 

Next, we discuss the conditions required to trigger the DTM in the Crab pulsar wind. 
The first step is the triggering of the simple tearing mode, which needs $k l \lesssim 0.6$ so that $l \lesssim \lambda/10$ where $\lambda = 2 \pi/k$ is the wavelength of the perturbation\footnote{Although the initial tearing instability needs somewhat long time to evolve in high-Lundquist number plasma, 
recent particle-in-cell simulations~\citep{2015MNRAS.448..606C,2015ApJ...801L..19P}  show that the current sheets in pulsar winds suffer from strong perturbation around light cylinder radius, 
and we expect such a strong perturbation helps the initial tearing instability to grow before reaching  $r = 50 r_{\rm L}$. }. The second step is that the DTM requires $k y_0 \lesssim 1$ such that $2\upi y_0 \lesssim \lambda$. 
In our case $y_0=3 l$ thus $6 \pi l \lesssim \lambda$ which is approximately comparable to the first condition. According to the pulse profile of the Crab, we have $l\approx 2 \pi r_{\rm L}/10$ therefore simple tearing is triggered for wavelengths $\lambda \gtrsim 2 \pi r_{\rm L}$ 
which will be easily satisfied since the sheet length in the striped wind is much longer than the sheet separation $2 \pi r_{\rm L}$.
The double tearing mode also demands the distance to a boundary from sheets should be larger than the sheet separation. Although the sheet separation in the striped wind is considered to be equal around the equator, in a high latitude region, it gradually becomes a combination of long separation $\lambda_{\rm l}$ and short separation $\lambda_{\rm s}$ which satisfies $\lambda_{\rm l} + \lambda_{\rm s} = 2 \pi r_{\rm L}$. This means the double tearing mode responsible for the GeV flare should occur around high-latitude region in the wind as indicated in Figure \ref{fig:4.3.1}. Since the inclination angle of Crab pulsar is considered to be larger than 45 degrees~\citep{1993ApJS...85..145R,1999ApJ...522.1046M}, we expect it is always possible to find a proper latitude in order for the growth of the double tearing mode. 

\begin{figure}
 \centering
  \includegraphics[width=8.cm,clip]{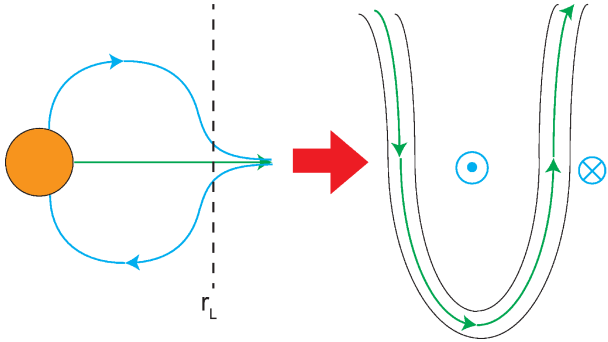}
  \caption{A schematic picture of generating guide field in the striped wind current sheets via a radial magnetic field component 
           in the pulsar magnetosphere. 
           The Green line is the radial magnetic field responsible for the guide field in the wind region. 
           The blue lines are the poloidal magnetic field responsible for the reconnection magnetic field in the wind region. 
          }
  \label{fig:4.3.2}
\end{figure}

Finally, as is indicated in Figure \ref{fig:3.2.3.1}, the double tearing mode becomes less explosive as the guide field increases. 
We consider the guide field in the current sheets will be small in the striped wind, 
and our scenario can still be applicable in the actual Crab pulsar wind. 
This is because, firstly, the radial magnetic field component decreases with increasing the radial coordinate as: $B_{\rm r} \propto (r/r_{\rm L})^{-2}$, 
so that it is negligible at $r \simeq 50 r_{\rm L}$; 
Secondly, the magnetic field in the striped wind is basically generated by the oscillation motion of the pulsar magnetosphere around the light cylinder 
where the magnetic field is theoretically considered to be anti-parallel across the current sheet in the equator. 
Hence, it is likely that the magnetic field in the wind region is also nearly anti-parallel across the sheets of striped wind. 
One possibility resulting in a strong guide field component in the current sheets is the existence of a monopole-like radial magnetic field in the sheet at the light cylinder. 
As indicated in Figure \ref{fig:4.3.2}, 
such a magnetic field will immediately change into a poloidal component in the sheet due to the oscillation motion, 
and survive globally in the striped wind region. 
However, the theoretical studies of pulsar magnetospheres do not support this monopole-like magnetic field structure, 
and this will be unlikely to occur. 
%
For these reasons, we expect the guide field to be very small, and does not reduce the energy release.

\section{Conclusions}
\label{sec:conclusions}

The striped wind structure represents a natural magnetic field configuration arising from the rotation of a magnetized neutron star surrounded by a relativistic pair plasma. It generates a current sheet wobbling around the equatorial plane. From simple analytical models, it is known that the magnetic field lines become mainly toroidal far from the neutron star with a poloidal component decreasing much faster than the toroidal component. Therefore, locally, the striped wind can be depicted as a current sheet with a guide field for which the strength depends on the distance from the light-cylinder.

Our new RRMHD simulations of the DTM including guide field effect confirm our previous works. DTM is a good candidate to explain the variability of the gamma-ray emission of the Crab flares. Indeed the explosive phase exhibits a magnetic reconnection event releasing magnetic energy into heat and therefore also into synchrotron radiation. The comparison between the synchrotron cooling time and the dynamical timescale shows that the plasma regime is far below the radiation reaction limit. The spectrum peaks at a energy~$\varepsilon_p$ simply related to the magnetization parameter~$\sigma$ by $\varepsilon_p\propto \sigma^2$. This allows us to constrain the magnetization of the Crab to lie around $\sigma\approx10^5$. This is in addition to the previous constraints on the location of the origin of the flares estimated to be around $r\approx50\,\rlight$ where $\rlight=c/\Omega$ is the light-cylinder radius, $c$ the speed of light and $\Omega$ the rotation speed of the pulsar and on the maximum Lorentz factor of the striped wind, $\Gamma\lesssim 300$ \citep{2015PPCF...57a4034P}.


\section*{Acknowledgments}
We would like to thank Yasunobu Uchiyama and Dmitry Khangulyan 
for many fruitful comments and discussions. 
Numerical computations were carried out on the Cray XC30 
at Center for Computational Astrophysics, CfCA, of National Astronomical Observatory of Japan.
Calculations were also carried out on SR16000 at YITP in Kyoto University. 
This work is supported in part by the Postdoctoral Fellowships for Research Abroad program by the Japan Society for the Promotion of Science No. 20130253 
and also by the Research Fellowship for Young Scientists (PD)
by the Japan Society for the Promotion of Science 
No.~20156571 (M.T.). 
J. P\'etri and H. Baty acknowledge financial support from the French National Research Agency (ANR) through the grant No. ANR-13-JS05-0003-01 (project EMPERE).


\appendix

\section{Convergence of the Calculations}
\label{sec:conv}

\begin{figure}
 \centering
  \includegraphics[width=8.cm,clip]{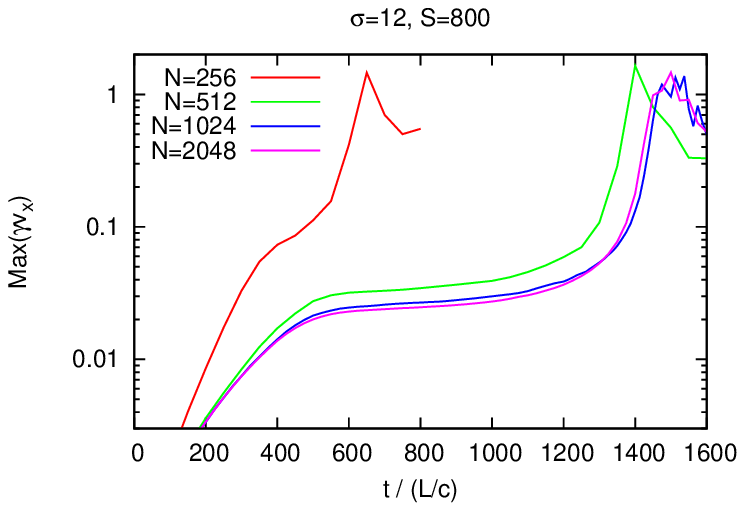}
  \caption{Temporal evolutions of the maximum 4-velocity of run RC2 using different numerical resolution. 
          }
  \label{fig:A.1}
\end{figure}

In this paper, we dealt with high-Lundquist number plasmas. 
In general, high resolution is necessary to simulate high-Lundquist number plasma to reduce numerical dissipation. 
Figure \ref{fig:A.1} shows temporal evolutions of the maximum velocity of run RC2 with different resolutions. 
It shows the resolution approximately $N_x \gtrsim S/2$ is sufficient for reproducing the maximum 4-velocity. 
Note that our code does not include viscosity and thermal conductivity but only resistivity, 
so that it is in general impossible to obtain perfect numerical convergence. 

\bsp

\label{lastpage}


\begin{thebibliography}{}

\bibitem[\protect\citeauthoryear{{Abdo} et~al.,}{{Abdo}
  et~al.}{2010}]{2010ApJ...708.1254A}
{Abdo} A.~A.,  et~al., 2010, ApJ, 708, 1254

\bibitem[\protect\citeauthoryear{{Baty}}{{Baty}}{2012}]{2012PhPl...19i2110B}
{Baty} H.,  2012, Physics of Plasmas, 19, 092110

\bibitem[\protect\citeauthoryear{{Baty}, {Petri} \& {Zenitani}}{{Baty}
  et~al.}{2013}]{2013MNRAS.436L..20B}
{Baty} H.,  {Petri} J.,    {Zenitani} S.,  2013, MNRAS, 436, L20

\bibitem[\protect\citeauthoryear{{Buehler} et~al.,}{{Buehler}
  et~al.}{2012}]{2012ApJ...749...26B}
{Buehler} R.,  et~al., 2012, ApJ, 749, 26

\bibitem[\protect\citeauthoryear{{B{\"u}hler} \& {Blandford}}{{B{\"u}hler} \&
  {Blandford}}{2014}]{2014RPPh...77f6901B}
{B{\"u}hler} R.,  {Blandford} R.,  2014, Reports on Progress in Physics, 77,
  066901

\bibitem[\protect\citeauthoryear{{Cerutti}, {Philippov}, {Parfrey} \&
  {Spitkovsky}}{{Cerutti} et~al.}{2015}]{2015MNRAS.448..606C}
{Cerutti} B.,  {Philippov} A.,  {Parfrey} K.,    {Spitkovsky} A.,  2015, MNRAS,
  448, 606

\bibitem[\protect\citeauthoryear{{Cerutti}, {Uzdensky} \& {Begelman}}{{Cerutti}
  et~al.}{2012}]{2012ApJ...746..148C}
{Cerutti} B.,  {Uzdensky} D.~A.,    {Begelman} M.~C.,  2012, ApJ, 746, 148

\bibitem[\protect\citeauthoryear{{Cerutti}, {Werner}, {Uzdensky} \&
  {Begelman}}{{Cerutti} et~al.}{2014}]{2014ApJ...782..104C}
{Cerutti} B.,  {Werner} G.~R.,  {Uzdensky} D.~A.,    {Begelman} M.~C.,  2014,
  ApJ, 782, 104

\bibitem[\protect\citeauthoryear{{Clausen-Brown} \& {Lyutikov}}{{Clausen-Brown}
  \& {Lyutikov}}{2012}]{2012MNRAS.426.1374C}
{Clausen-Brown} E.,  {Lyutikov} M.,  2012, MNRAS, 426, 1374

\bibitem[\protect\citeauthoryear{{Coroniti}}{{Coroniti}}{1990}]{1990ApJ...349..538C}
{Coroniti} F.~V.,  1990, ApJ, 349, 538

\bibitem[\protect\citeauthoryear{{Evans} \& {Hawley}}{{Evans} \&
  {Hawley}}{1988}]{1988ApJ...332..659E}
{Evans} C.~R.,  {Hawley} J.~F.,  1988, ApJ, 332, 659

\bibitem[\protect\citeauthoryear{{Kirk}, {Lyubarsky} \& {Petri}}{{Kirk}
  et~al.}{2009}]{2009ASSL..357..421K}
{Kirk} J.~G.,  {Lyubarsky} Y.,    {Petri} J.,  2009, in {Becker} W.,  ed.,
  Astrophysics and Space Science Library Vol.~357 of Astrophysics and Space
  Science Library, {The Theory of Pulsar Winds and Nebulae}.
p.~421

\bibitem[\protect\citeauthoryear{{Kirk}, {Skj{\ae}raasen} \& {Gallant}}{{Kirk}
  et~al.}{2002}]{2002A&A...388L..29K}
{Kirk} J.~G.,  {Skj{\ae}raasen} O.,    {Gallant} Y.~A.,  2002, AAP, 388, L29

\bibitem[\protect\citeauthoryear{{Komissarov}, {Barkov} \&
  {Lyutikov}}{{Komissarov} et~al.}{2007}]{2007MNRAS.374..415K}
{Komissarov} S.~S.,  {Barkov} M.,    {Lyutikov} M.,  2007, MNRAS, 374, 415

\bibitem[\protect\citeauthoryear{{Lyubarsky}}{{Lyubarsky}}{2005}]{2005MNRAS.358..113L}
{Lyubarsky} Y.~E.,  2005, MNRAS, 358, 113

\bibitem[\protect\citeauthoryear{{Lyutikov} \& {Uzdensky}}{{Lyutikov} \&
  {Uzdensky}}{2003}]{2003ApJ...589..893L}
{Lyutikov} M.,  {Uzdensky} D.,  2003, ApJ, 589, 893

\bibitem[\protect\citeauthoryear{{Michel}}{{Michel}}{1971}]{1971CoASP...3...80M}
{Michel} F.~C.,  1971, Comments on Astrophysics and Space Physics, 3, 80

\bibitem[\protect\citeauthoryear{{Moffett} \& {Hankins}}{{Moffett} \&
  {Hankins}}{1999}]{1999ApJ...522.1046M}
{Moffett} D.~A.,  {Hankins} T.~H.,  1999, ApJ, 522, 1046

\bibitem[\protect\citeauthoryear{{Palmer} et~al.,}{{Palmer}
  et~al.}{2005}]{2005Natur.434.1107P}
{Palmer} D.~M.,  et~al., 2005, Nature, 434, 1107

\bibitem[\protect\citeauthoryear{{P{\'e}tri}}{{P{\'e}tri}}{2012}]{2012MNRAS.424.2023P}
{P{\'e}tri} J.,  2012, MNRAS, 424, 2023

\bibitem[\protect\citeauthoryear{{P{\'e}tri}}{{P{\'e}tri}}{2013}]{2013MNRAS.434.2636P}
{P{\'e}tri} J.,  2013, MNRAS, 434, 2636

\bibitem[\protect\citeauthoryear{{P{\'e}tri} \& {Kirk}}{{P{\'e}tri} \&
  {Kirk}}{2005}]{2005ApJ...627L..37P}
{P{\'e}tri} J.,  {Kirk} J.~G.,  2005, ApJL, 627, L37

\bibitem[\protect\citeauthoryear{{P{\'e}tri}, {Takamoto}, {Baty} \&
  {Zenitani}}{{P{\'e}tri} et~al.}{2015}]{2015PPCF...57a4034P}
{P{\'e}tri} J.,  {Takamoto} M.,  {Baty} H.,    {Zenitani} S.,  2015, Plasma
  Physics and Controlled Fusion, 57, 014034

\bibitem[\protect\citeauthoryear{{Philippov}, {Spitkovsky} \&
  {Cerutti}}{{Philippov} et~al.}{2015}]{2015ApJ...801L..19P}
{Philippov} A.~A.,  {Spitkovsky} A.,    {Cerutti} B.,  2015, ApJ, 801, L19

\bibitem[\protect\citeauthoryear{{Rankin}}{{Rankin}}{1993}]{1993ApJS...85..145R}
{Rankin} J.~M.,  1993, ApJS, 85, 145

\bibitem[\protect\citeauthoryear{{Somov} \& {Verneta}}{{Somov} \&
  {Verneta}}{1993}]{1993SSRv...65..253S}
{Somov} B.~V.,  {Verneta} A.~I.,  1993, SSR, 65, 253

\bibitem[\protect\citeauthoryear{{Striani} et~al.,}{{Striani}
  et~al.}{2011}]{2011ApJ...741L...5S}
{Striani} E.,  et~al., 2011, ApJL, 741, L5

\bibitem[\protect\citeauthoryear{{Striani} et~al.,}{{Striani}
  et~al.}{2013}]{2013ApJ...765...52S}
{Striani} E.,  et~al., 2013, ApJ, 765, 52

\bibitem[\protect\citeauthoryear{{Takahashi}, {Kudoh}, {Masada} \&
  {Matsumoto}}{{Takahashi} et~al.}{2011}]{2011ApJ...739L..53T}
{Takahashi} H.~R.,  {Kudoh} T.,  {Masada} Y.,    {Matsumoto} J.,  2011, ApJ,
  739, L53

\bibitem[\protect\citeauthoryear{{Takamoto}}{{Takamoto}}{2013}]{2013ApJ...775...50T}
{Takamoto} M.,  2013, ApJ, 775, 50

\bibitem[\protect\citeauthoryear{{Takamoto} \& {Inoue}}{{Takamoto} \&
  {Inoue}}{2011}]{2011ApJ...735..113T}
{Takamoto} M.,  {Inoue} T.,  2011, ApJ, 735, 113

\bibitem[\protect\citeauthoryear{{Takamoto}, {Kisaka}, {Suzuki} \&
  {Terasawa}}{{Takamoto} et~al.}{2014}]{2014ApJ...787...84T}
{Takamoto} M.,  {Kisaka} S.,  {Suzuki} T.~K.,    {Terasawa} T.,  2014, ApJ,
  787, 84

\bibitem[\protect\citeauthoryear{{Uzdensky}, {Cerutti} \&
  {Begelman}}{{Uzdensky} et~al.}{2011}]{2011ApJ...737L..40U}
{Uzdensky} D.~A.,  {Cerutti} B.,    {Begelman} M.~C.,  2011, ApJL, 737, L40

\bibitem[\protect\citeauthoryear{{Zanotti} \& {Dumbser}}{{Zanotti} \&
  {Dumbser}}{2011}]{2011MNRAS.418.1004Z}
{Zanotti} O.,  {Dumbser} M.,  2011, MNRAS, 418, 1004

\bibitem[\protect\citeauthoryear{{Zenitani}, {Hesse} \& {Klimas}}{{Zenitani}
  et~al.}{2009}]{2009ApJ...705..907Z}
{Zenitani} S.,  {Hesse} M.,    {Klimas} A.,  2009, ApJ, 705, 907

\end{thebibliography}
\end{document}